\documentclass[prc,aps,twocolumn,showpacs,nofootinbib]{revtex4-1}
\usepackage{}
\usepackage{amsmath}
\usepackage{amssymb}
\usepackage{amsthm}
\usepackage{amsfonts}
\usepackage{graphicx}
\usepackage[usenames,dvipsnames]{color}
\usepackage{epstopdf}
\usepackage{afterpage}
\usepackage{mathrsfs}
\usepackage{graphicx,latexsym,epsfig}
\usepackage{mathrsfs}
\usepackage{color}
\usepackage{bm}
\usepackage{CJK}
\usepackage[colorlinks,linkcolor=red,anchorcolor=blue,citecolor=blue]{hyperref}

\newcommand{\beq}{\vspace{0.5em}\begin{equation}}
\newcommand{\eeq}{\end{equation}\vspace{0.5em}}
\newcommand{\beqn}{\vspace{0.5em}\begin{eqnarray}}
\newcommand{\eeqn}{\end{eqnarray}\par\vspace{0.5em}\noindent}

\newcommand{\bsub}{\begin{subequations}}
\newcommand{\esub}{\end{subequations}}

\begin{document}

\title{Quadrupole collectivity and  shell closure in neutron-rich nuclei around  $N=126$}

 \author{X. Y. Wu}
  \affiliation{College of Physics and Communication Electronics, Jiangxi Normal University, Nanchang 330022, China}
 \author{J. M. Yao}\email{Corresponding author: yaoj@frib.msu.edu}
 \affiliation{FRIB/NSCL Laboratory, Michigan State University, East Lansing, Michigan 48824, USA}
 \date{\today}

%
\begin{abstract}

 We present a comprehensive study on the low-lying states of neutron-rich Er, Yb, Hf, and W isotopes across the  $N=126$ shell with a multi-reference covariant density functional theory.  Beyond mean-field effects from shape mixing and symmetry restoration on the observables that are relevant for understanding  quadrupole collectivity and underlying shell structure are investigated. The general features of low-lying states in closed-shell nuclei are retained  in these four isotopes around $N=126$,  even though the shell gap is overall quenched by about 30\% with the beyond mean-field effects. These effects are consistent with the previous generator-coordinate calculations based on Gogny forces, but much smaller than that predicted by the collective Hamiltonian calculation.  It implies that the beyond mean-field effects on the  $r$-process abundances before the third peak at $A\sim195$ might be more moderate than that found in A. Arcones and G. F. Bertsch, Phys. Rev. Lett. 108, 151101 (2012).

 \end{abstract}
\maketitle


\section{\label{introduction}Introduction}

The knowledge of neutron-rich nuclei far away from the $\beta$-stability line is essential to understand nucleosynthesis and the origin of heavy nuclei in the Universe.
About half of the  elements with mass number $A\geqslant60$ are produced within the rapid neutron-capture process ($r$-process)~\cite{Burbidge57,Cowan91,Arnould07}.
 Under astrophysical environments with extreme neutron densities, neutron captures are much faster than $\beta$ decays, and the $r$-process path runs
through nuclei with large neutron excess. Along the path, presence of shell closure and large shell gap determines where the material accumulates.
For example,  the peaks around  $A = 80, 130$,  and 195 in the $r$-process abundances are  mainly attributed to the neutron $N=50, 82$, and 126 shell closures, respectively.
Previous studies have shown that quenching of the $N = 82$ shell gap has a substantial influence on the predicted abundances~\cite{Chen95,Pfeiffer97}, even though there is a controversial on whether
or not it should be quenched~\cite{Kautzsch00,Dillmann03,Jungclaus07,Watanabe13}.  Experimental studies of the neutron-rich nuclei across the $N=126$  shell are more challenging as the production of these nuclei from the reactions of nuclear fusion, fission, and fragmentation is very low.  Therefore, knowledge on the $N=126$ shell gap in neutron-rich nuclei relies heavily on nuclear model predictions.
 
Nuclear density functional theory  (DFT) starting from a universal energy  functional with about a dozen parameters fitted to a set of nuclear properties provides currently  the only microscopic tool to study neutron-rich nuclei across the $N=126$ shell.  The information on the  shell structure can be learnt either from nucleon separation energies or  from the systematics in  low-energy spectroscopic quantities. A strong shell quenching  in an even-even nucleus  is usually signaled by  a moderate change in two-nucleon separation energies and by an enhancement in low-lying collective excitations.
On the mean-field level, the predicted two-neutron separation energies  with different energy  functionals differ from each other by a factor of about  two for the neutron-rich nuclei around $N=126$~\cite{Lalazissis99,Stoitsov03, Bertsch05, Geng05a,Erler12,Afanasjev16,Xia18}.  This discrepancy contributes largely to the uncertainty in the predicted $r$-process abundances before the third peak at $A\sim195$.
In the recent decade, beyond mean-field (BMF) effects from symmetry restoration and configuration mixing on the predicted nuclear masses and nucleon separation energies
have been investigated within the framework of generator coordinate method (GCM) with either Skyrme or Gogny energy functionals~\cite{Bender05,Bender06,Bender08,Tomas15,Rodriguez15}.  Generally speaking, the inclusion of the BMF effects leads to a quenching of the shell gap and
is shown to improve the description of  two-nucleon separation energies in known nuclei around shell closure. A similar quenching is also predicted  
for the $N = 126$ shell gap in neutron-rich region~\cite{Tomas15,Rodriguez15}.

The covariant formulation of DFT (CDFT) has achieved a comparable success in many aspects of applications to nuclear physics~\cite{Vretenar05,Meng06,Meng16}. In particular, the spin-orbit interaction of nucleons emerges automatically in the relativistic framework. This 
feature is important for understanding nuclear shell structure in neutron-rich nuclei.  Considering these facts, it is  interesting to revisit the low-lying states of neutron-rich nuclei across the  $N=126$ shell  within this framework, shedding light on the $N=126$ shell gap from different perspectives.  With the CDFT, the BMF effects  associated with rotational motion and quadrupole shape vibrational motion for 575 even-even nuclei with proton numbers ranging from $Z=8$ to $Z=108$ have been evaluated using either cranking approximation~\cite{Zhang14} or five-dimensional collective Hamiltonian (5DCH)~\cite{Lu15}. It was shown that both the masses and two-neutron separation energies are significantly improved. A more accurate evaluation of the BMF effects on these quantities requires more computationally expensive calculations using quantum-number projected GCM. With recent extensions to the multi-reference framework~\cite{Yao09,Yao10,Yao14}, it is feasible to carry out such kind of studies. In this work, we are focused on the low-lying states of neutron-rich Er, Yb, Hf, and W isotopes with neutron numbers ranging between $122 \leqslant N \leqslant138$.  We compare our results with the predictions by other models on either mean-field or beyond mean-field level.

The paper is arranged as follows. In Sec.~\ref{Sec.II},  a brief introduction to the theoretical framework is presented. This includes both the CDFT and its extension to multi-reference version with projection and  GCM. The numerical details are given in Sec.~\ref{Sec.III}. Results  are analyzed in Sec.~\ref{Sec.IV}. Finally, a summary and outline are provided in Sec.~\ref{Sec.V}.


\section{Theoretical framework}
 \label{Sec.II}

 
 \subsection{Covariant density functional theory} 
  \label{subSec.IIA}


Starting from a nonlinear point-coupling effective Lagrangian and taking mean-field approximation, one finds the energy of nuclear systems as a  function of local  densities and currents \cite{Burvenich02,Zhao10}
 \begin{eqnarray}
 {{E}}_{\rm RMF}
 &=&\int d\bm{r} \sum_{k=1}^A{~v_k^2
 ~{\bar{\psi}_k (\bm{r}) \left( -i\bm{\gamma}\bm{\nabla} + m\right )\psi_k(\bm{r})}}\nonumber \\
 &+& \int d{\bm r }~{\left(\frac{\alpha_S}{2}\rho_S^2+\frac{\beta_S}{3}\rho_S^3 +
 \frac{\gamma_S}{4}\rho_S^4+\frac{\delta_S}{2}\rho_S\triangle \rho_S \right.}\nonumber \\
 &+&  {\left.\frac{\alpha_V}{2}j_\mu j^\mu + \frac{\gamma_V}{4}(j_\mu j^\mu)^2 +
 \frac{\delta_V}{2}j_\mu\triangle j^\mu \right.} \nonumber \\
 &+& \left. \frac{\alpha_{TV}}{2}j^{\mu}_{TV}(j_{TV})_\mu+\frac{\delta_{TV}}{2}
 j^\mu_{TV}\triangle  (j_{TV})_{\mu}\right.\nonumber \\
 &+&\frac{\alpha_{TS}}{2}\rho_{TS}^2
 \left.+\frac{\delta_{TS}}{2}\rho_{TS}\triangle
 \rho_{TS} +ej^\mu_p A_\mu
 \right) ,
 \label{EDF}
 \end{eqnarray}
 where $\psi_k(\bm{r})$ is a Dirac spinor for the single-nucleon wave function.  The coupling constants $\alpha_i, \beta_i, \gamma_i$, and $\delta_i$ are  free parameters.
 $j^\mu_p$ is the current of protons and $A_\mu$ represents the electromagnetic field.  The four types of densities or currents: isoscalar-scalar ($S$),  isovector-scalar ($TS$),  isoscalar-vector ($V$) and isovector-vector ($TV$) are defined as
  \begin{subequations}
 \begin{align}
 \label{dens_1}
 \rho_{S}({\bm r}) &=\sum_{k} v_k^2 ~\bar{\psi}_{k}({\bm r})
              \psi _{k}({\bm r})~,  \\
 \label{dens_2}
 \rho_{TS}({\bm r}) &=\sum_{k} v_k^2 ~
       \bar{\psi}_{k}({\bm r})\tau_3\psi _{k}^{{}}({\bm r})~,  \\
 \label{dens_3}
 j^{\mu}({\bm r}) &=\sum_{k}  v_k^2 ~\bar{\psi}_{k}({\bm r})
         \gamma^\mu\psi _{k}^{{}}({\bm r})~,  \\
 \label{dens_4}
 j^{\mu}_{TV}({\bm r}) &=\sum_{k}  v_k^2 ~\bar{\psi}_{k}({\bm r})
      \gamma^\mu \tau_3 \psi _{k}^{{}}({\bm r}).
       \end{align}
 \end{subequations}
The summation in  Eqs.~(\ref{dens_1}) - (\ref{dens_4}) runs over all occupied states in the Fermi sea with nonzero occupation probability $v_k^2$, which is determined
 with the BCS approximation.

 The single-nucleon wave function $\psi_{k}({\bm r})$  is determined by the following Dirac equation
 \begin{eqnarray}
 \label{Dirac:N}
 \left\{\bm{\alpha}\cdot\bm{p}+V^0({\bm r}) +\beta \left[m+S({\bm r})\right]\right\}\psi_{k}({\bm r})=\epsilon_k\psi_{k}({\bm r})\, ,
 \end{eqnarray}
 where the scalar and vector potentials
 \begin{subequations}\begin{eqnarray}
 \label{scapot}
 S({\bm r}) & = & \Sigma_S({\bm r}) + \tau_3\Sigma_{TS}({\bm r})\; ,\\
 \label{vecpot}
 V^{\mu}({\bm r}) & = &\Sigma^{\mu}_V({\bm r}) + \tau_3\Sigma^{\mu}_{TV}({\bm r})\; ,
 \end{eqnarray}\end{subequations}
 contain nucleon isoscalar-scalar, isovector-scalar,
 isoscalar-vector and isovector-vector self-energies as follows
 \begin{subequations}\begin{eqnarray}
    \label{selfS}
    \Sigma_S & = & \alpha_S \rho_S + \beta_S \rho_S^2 +
    \gamma_S\rho_S^3+ \delta_S \triangle \rho_S \; ,\\
    \label{selfTS}
    \Sigma_{TS} & = & \alpha_{TS} \rho_{TS}
    +\delta_{TS} \triangle \rho_{TS}\; , \\
    \label{selfV}
    \Sigma^{\mu}_V& = & \alpha_V j^{\mu}
   +\gamma_V (j_\nu j^\nu)j^\mu + \delta_V \triangle j^\mu \nonumber \\
    &&-eA^\mu\frac{1-\tau_3}{2} \; ,\\
    \label{selfTV}
    \Sigma^{\mu}_{TV} & = & \alpha_{TV} j^{\mu}_{TV}
       + \delta_{TV} \triangle j^{\mu}_{TV}.
 \end{eqnarray}
 \end{subequations}
 
 For even-even nuclei, only the zero-component of the vector potentials is nonzero in Eq. (\ref{Dirac:N}).  
 Besides, in order to generate a set of mean-field solutions with different intrinsic deformation,   a quadratic constraint term on the mass quadrupole moment is added onto the energy in the variational calculation,
 \begin{equation}
 \dfrac{\delta }{\delta\bar\psi_k}\left[E_{\rm RMF} + \sum_{\mu=0, 2} C_{2\mu}(\langle \hat{Q}_{2\mu}\rangle - q_{2\mu})^2  \right]=0,
 \label{constraint}
 \end{equation}
 which generates a constrained potential term to Eq. (\ref{Dirac:N}).  The $C_{2\mu}$ is a stiffness parameter and $\langle \hat{Q}_{2\mu}\rangle$ denotes the expectation value of the mass quadrupole moment operator
 \begin{subequations}\begin{eqnarray}
  \!\!\langle\hat{Q}_{20}\rangle \!\!& = & \!\! \sqrt{\dfrac{5}{16\pi}} \langle 2z^2 - x^2 - y^2\rangle=\frac{3}{4\pi}AR_0^2\beta\cos\gamma\, ,\\
  \!\!\langle\hat{Q}_{22}\rangle \!\!& = & \!\!\sqrt{\dfrac{15}{32\pi}} \langle x^2 - y^2\rangle=\frac{3}{4\pi}AR_0^2\frac{1}{\sqrt 2}\beta\sin\gamma\, .
 \end{eqnarray}\end{subequations}
 
 For the sake of simplicity, we constrain the parameter $\gamma$ to be either 180 degree or zero degree, which corresponds to the nucleus with either oblate or prolate deformation, keeping the $z$-axis always being the symmetric axis. With this simplification, only one-dimensional angular momentum projection will be needed to restore rotational symmetry. 
The deformation parameter $\beta$ is calculated by $\beta= \dfrac{4\pi}{3AR_0^2} {\langle \hat Q_{20}\rangle}$, where $R_0=1.2A^{1/3}$ (fm), and $A$ is the mass number.
 

  \subsection{Beyond mean-field approximation with generator coordinate method}
   \label{subSec.IIB}

The collective wave function of low-lying states is constructed as a linear combination of particle-number and angular-momentum projected mean-field wave functions
 \beqn%
 \label{wavefun}
 \vert \Psi^{JM}_\alpha\rangle
 =\sum_\beta f^{J}_\alpha(\beta)\hat P^J_{MK=0} \hat P^N\hat P^Z\vert \Phi(\beta)\rangle.
 \eeqn
 where the intrinsic shape of the mean-field wave function is restricted to have axial symmetry.
 The $\alpha$ labels different collective states for a given angular momentum $J$.
 The $\hat P^N$,  $\hat P^Z$, and $\hat P^J_{M0}$  are projection operators onto good quantum numbers,
 i.e., neutron number $N$, proton number $Z$, and angular momentum $J$, respectively.
 The weight functions $f^{J}_\alpha(\beta)$ are determined by minimizing the energy of the collective state with
 respect to the weight function. This leads to the Hill-Wheeler-Griffin (HWG) equation~\cite{Hill53,Griffin57,Ring80}
 \beqn%
 \label{HWGE}
 \sum_{\beta^\prime} \left[ \mathcal{H}^{J}(\beta,\beta^\prime)
          - E_\alpha^{J} \, \mathcal{N}^{J}(\beta,\beta^\prime)
           \right] f_\alpha^{J}(\beta^\prime) = 0 \, .
 \eeqn%
 where the norm kernel
${\cal N}^J(\beta,\beta')$
and the Hamiltonian kernel
${\cal H}^J(\beta,\beta')$
are defined as
\beq
{\cal O}^J(\beta,\beta')\equiv
 \langle \Phi(\beta) \vert \hat O
\hat{P}^J_{00}\hat{P}^N \hat{P}^Z \vert\Phi
(\beta')\rangle,
\eeq
with $\hat{O}=1$ and $\hat{O}=\hat{H}$, respectively. The solution of the Eq.~(\ref{HWGE})
 provides the weight functions $f^{J}_\alpha(\beta)$ and the energy spectrum, as well as other information needed for calculating
 the electric multipole transition strengths. This framework is called multi-reference covariant density functional theory (MR-CDFT).
 More details on the framework could be found in  Ref.~\cite{Yao10}.

 The electric quadrupole transition strength $B(E2)$ from the initial state
 $(J_i,\alpha_i)$ to the final state $(J_f,\alpha_f)$ is calculated as follows
 \begin{eqnarray}
  &&B(E2; J_i,\alpha_i\rightarrow J_f,\alpha_f)\nonumber\\
    &=& \frac{1}{2J_i+1}
        \Big\vert\sum_{\beta^\prime,\beta}f_{\alpha_f}^{J_{f}\ast}(\beta^\prime)  \langle J_f,\beta^\prime\vert\vert \hat Q_{2}\vert\vert J_i,\beta\rangle
        f_{\alpha_i}^{J_{i}}(\beta)\Big\vert^2\, ,\nonumber\\
 \label{BE2}
 \end{eqnarray}
 with the reduced transition matrix element
 \begin{eqnarray}
 &&\langle J_f,\beta^\prime\vert\vert \hat Q_{2}\vert\vert J_i,\beta\rangle\nonumber\\
 &=&\frac{(2J_f+1)(2J_i+1)}{2}
 \sum_{M=-2}^{+2}
 \left(\begin{array}{ccc}
   J_f  &  2     &J_i \\
   0    & M      &-M \\
 \end{array}
 \right) \nonumber\\
 &&  \int_0^\pi d\theta\,\sin(\theta)\,d_{-M0}^{J_i\ast}(\theta)\langle\Phi(\beta^\prime)
 \vert\hat Q_{2M}e^{i\theta\hat J_y}\hat P^N \hat P^Z\vert\Phi(\beta)\rangle \, ,\nonumber\\
  \label{Q2}
 \end{eqnarray}
 where $\hat Q_{2M}\equiv er^2Y_{2M}$ is the electric quadrupole moment operator.
 Meanwhile, we can also calculate the spectroscopic quadrupole moment for each state
 \begin{eqnarray}\label{qspec}
 Q^{\rm spec}(J_\alpha^\pi)
 &=& \sqrt{\frac{16\pi}{5}}
 \left(\begin{array}{ccc}
    J  &  2         &J \\
    J & 0      &-J \\
 \end{array}
 \right)\sum_{\beta,\beta^\prime}f_\alpha^{J\ast}(\beta^\prime) \nonumber\\
 &&\times \langle J,\beta^\prime
 \vert\vert\hat Q_2\vert\vert J,\beta\rangle f_\alpha^{J}(\beta) \, .
 \end{eqnarray}

 Since the $B(E2)$ values and spectroscopic quadrupole moments $Q^{\texttt{spec}}(J_\alpha^\pi)$ are
 calculated in the full configuration space, there is no need to introduce effective charge,
 and $e$ simply corresponds to bare value of the proton charge.

 \section{Numerical details}
 \label{Sec.III}

 Parity, time-reversal invariance and axial symmetry are assumed. The Dirac equation (\ref{Dirac:N}) is solved by expanding the Dirac spinor in terms of three-dimensional harmonic oscillator basis in Cartesian coordinate with 14 major shells which are found to be sufficient for the nuclei under consideration. The relativistic energy density functional PC-PK1~\cite{Zhao10} is employed in the calculations. Pairing correlations between nucleons are treated with the BCS approximation using a density-independent $\delta$ force with a smooth cutoff factor~\cite{Krieger90}. The Gauss-Legendre quadrature is used for the integrals over the Euler angle $\theta$ in the calculations of the projected kernels. The number of mesh points in the interval $\theta\in[0,\pi]$ for the $\theta$ is chosen as $N_\theta=14$. The number of mesh points for the gauge angles in the Fomenko's expansion~\cite{Fomenko70} for the particle-number projection is $N_\phi=9$. The Pfaffian method~\cite{Robledo09} has been implemented to calculate the norm overlap in the kernels.

 \section{Results and discussions}
 \label{Sec.IV}

 \subsection{Analysis of quadrupole collectivity}

 \begin{figure*}[]
 \centering
 \includegraphics[width=18.0cm]{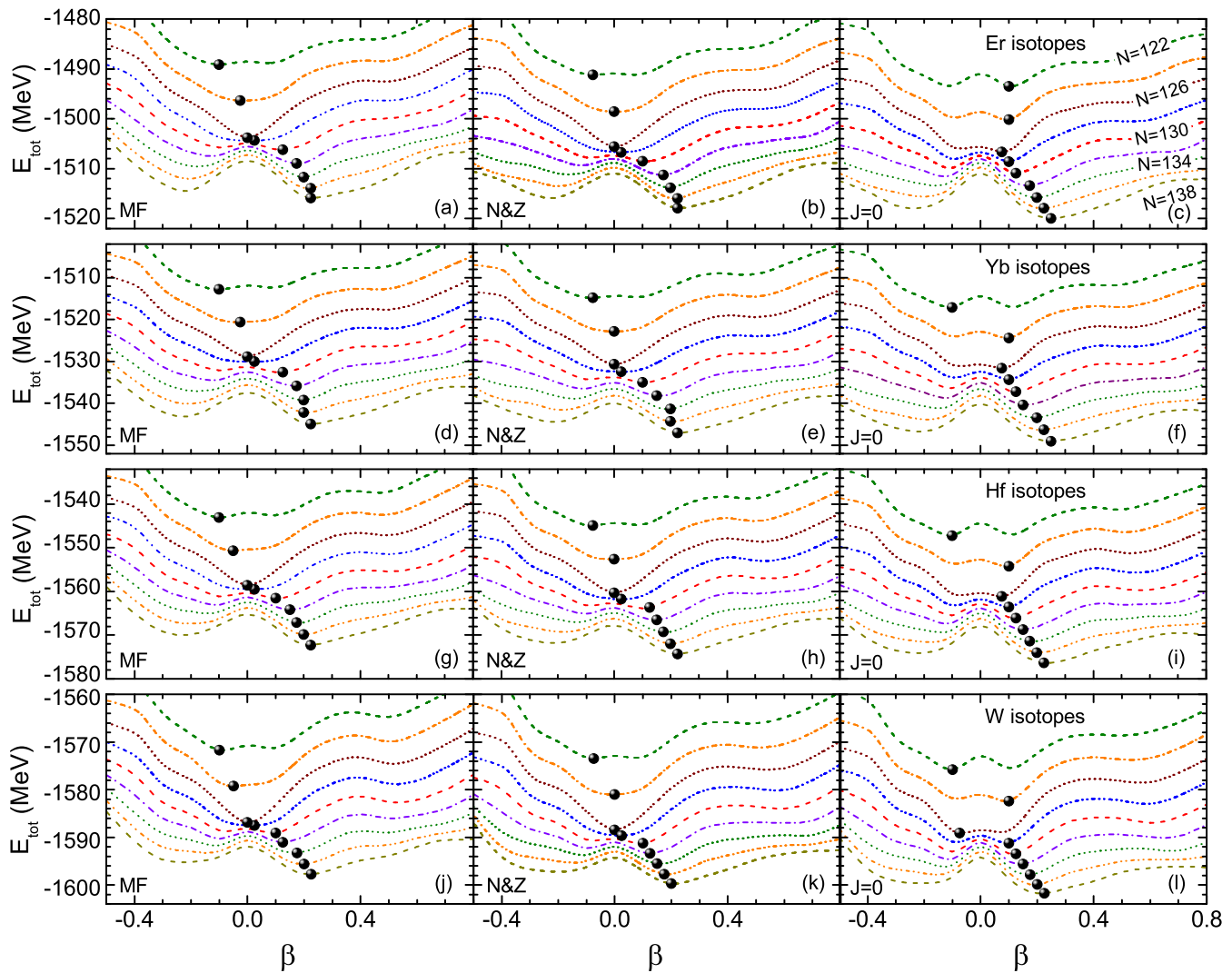}
 \caption{(Color online) Energies of mean-field states (MF, left),  particle-number projected states ($N\&Z$, middle),
           and those with additional projection onto angular momentum ($J=0$, right) for (a-c) $^{190-206}$Er, (d-f) $^{192-208}$Yb, (g-i) $^{194-210}$Hf, and (j-l) $^{196-212}$W isotopes as a function of the intrinsic mass quadrupole deformation. The energies of $^{190-206}$Er, $^{192-208}$Yb, $^{194-210}$Hf, and $^{196-212}$W isotopes are shifted by $-4, -3, -2,$ and $-2$ MeV  between two neighboring isotopes, respectively.
The global energy minima are indicated by black dots. }
 \label{fig:PECs_all}
 \end{figure*}

\begin{figure}[h]
 \centering
 \includegraphics[width=8.5cm]{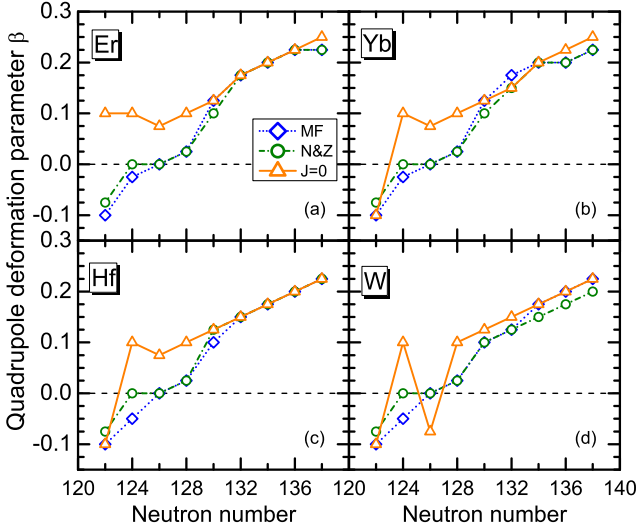}
 \caption{(Color online) The quadrupole deformation parameter $\beta$ of ground state obtained by MF, PNP ($N\&Z$), and PNAMP ($J=0$) calculations using PC-PK1 force, as a function of neutron number for $^{190-206}$Er, $^{192-208}$Yb, $^{194-210}$Hf, and $^{196-212}$W isotopes.}
 \label{fig:beta_all}
 \end{figure}

Figure~\ref{fig:PECs_all} displays the energy surfaces for the even-even neutron-rich Er, Yb, Hf, and W isotopes from  both mean-field and beyond mean-field calculations. It  shows evidently the development of quadrupole collectivity globally with the increase of neutron number from $N=126$. The quadrupole deformation parameter $\beta$ at each global energy minimum  is displayed in Fig.~\ref{fig:beta_all} as a function of neutron number. Along each isotopic chain, the equilibrium quadrupole shape undergoes a transition from oblate shape to prolate one while across the neutron number $N=126$ with spherical shape. The angular-momentum projection brings an additional energy to the weakly deformed configurations and thus changes somewhat the location of the energy minimum in the nuclei around $N=126$.  For those weakly deformed nuclei and the transitional nuclei, the concept of nuclear shape is ill-defined as  a large shape mixing effect is expected there. After mixing differently-shaped configurations, one ends up with a more correlated wave function for nuclear ground state. The low-lying excited states associated with rotational and vibrational collective excitations are also obtained. The properties of these low-lying states are used to provide information on the underlying shell structure.

\begin{figure}[]
 \centering
 \includegraphics[width=8.5cm]{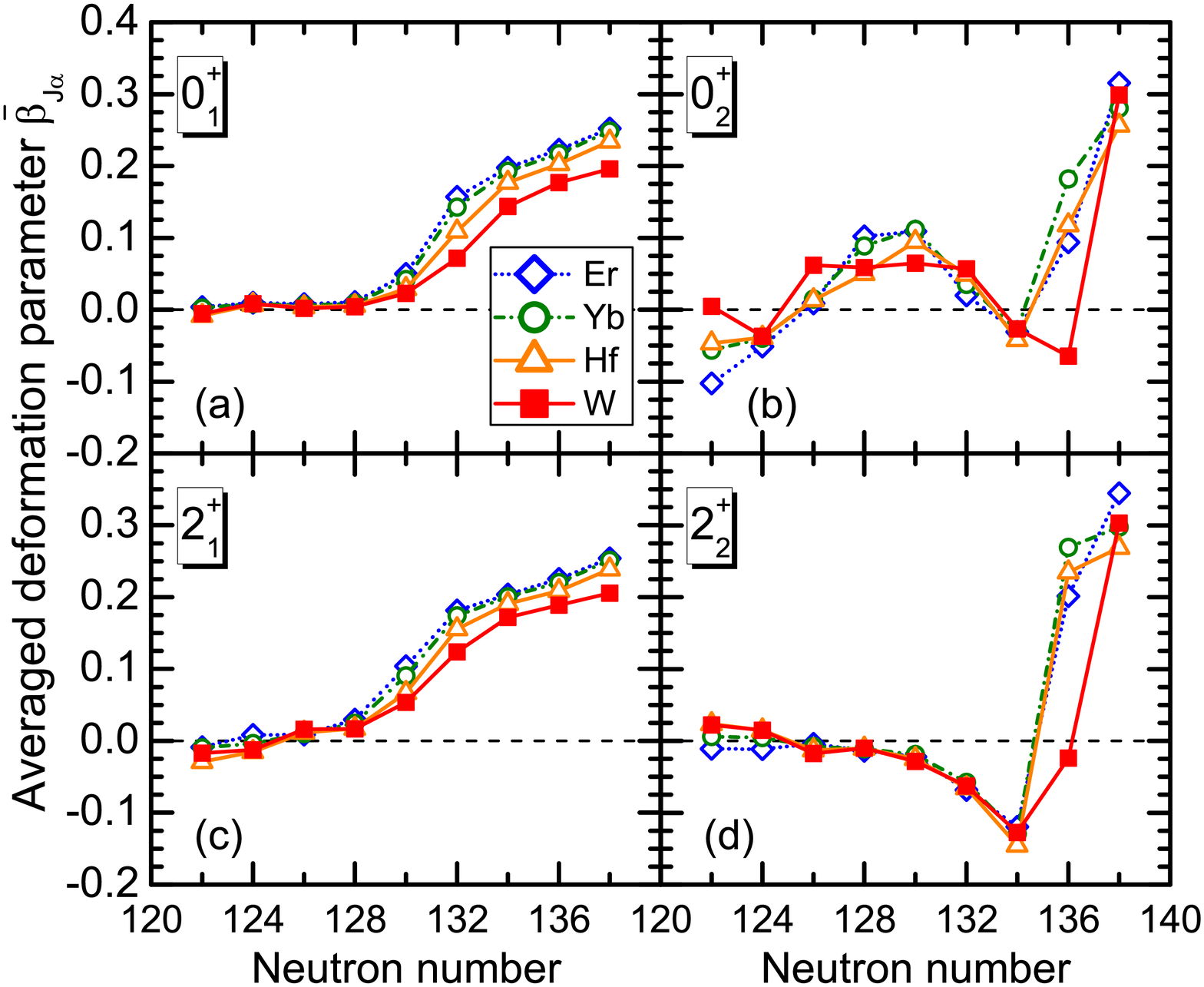}
 \caption{(Color online) The averaged deformation parameter $\bar\beta_{J\alpha}$ for the first two  $J^\pi = 0^+$ and $2^+$ states in $^{190-206}$Er, $^{192-208}$Yb, $^{194-210}$Hf,  and $^{196-212}$W isotopes as a function of neutron number, respectively.  }
 \label{fig:gs_beta_all}
 \end{figure}

 Figure~\ref{fig:gs_beta_all} displays the averaged  deformation parameter $\bar\beta_{J\alpha}$ for the first  two $0^+$ and $2^+$ states, which is defined as
  \beqn
 \bar\beta_{J\alpha}=\sum_{\beta} \beta \vert g^{J}_\alpha(\beta)\vert^2,
  \eeqn
 where the collective wave function  $g_\alpha^J(\beta)$  is related to the mixing weight~\cite{Ring80}
 \beqn%
 \label{gfunc}
 g_\alpha^{J}(\beta)=\sum_{\beta^\prime}\left[\mathcal{N}^{J}
 \right]^{1/2}(\beta,\beta^\prime) f_\alpha^{J}(\beta^\prime).
 \eeqn%

\begin{figure}[]
 \centering
 \includegraphics[width=8.5cm]{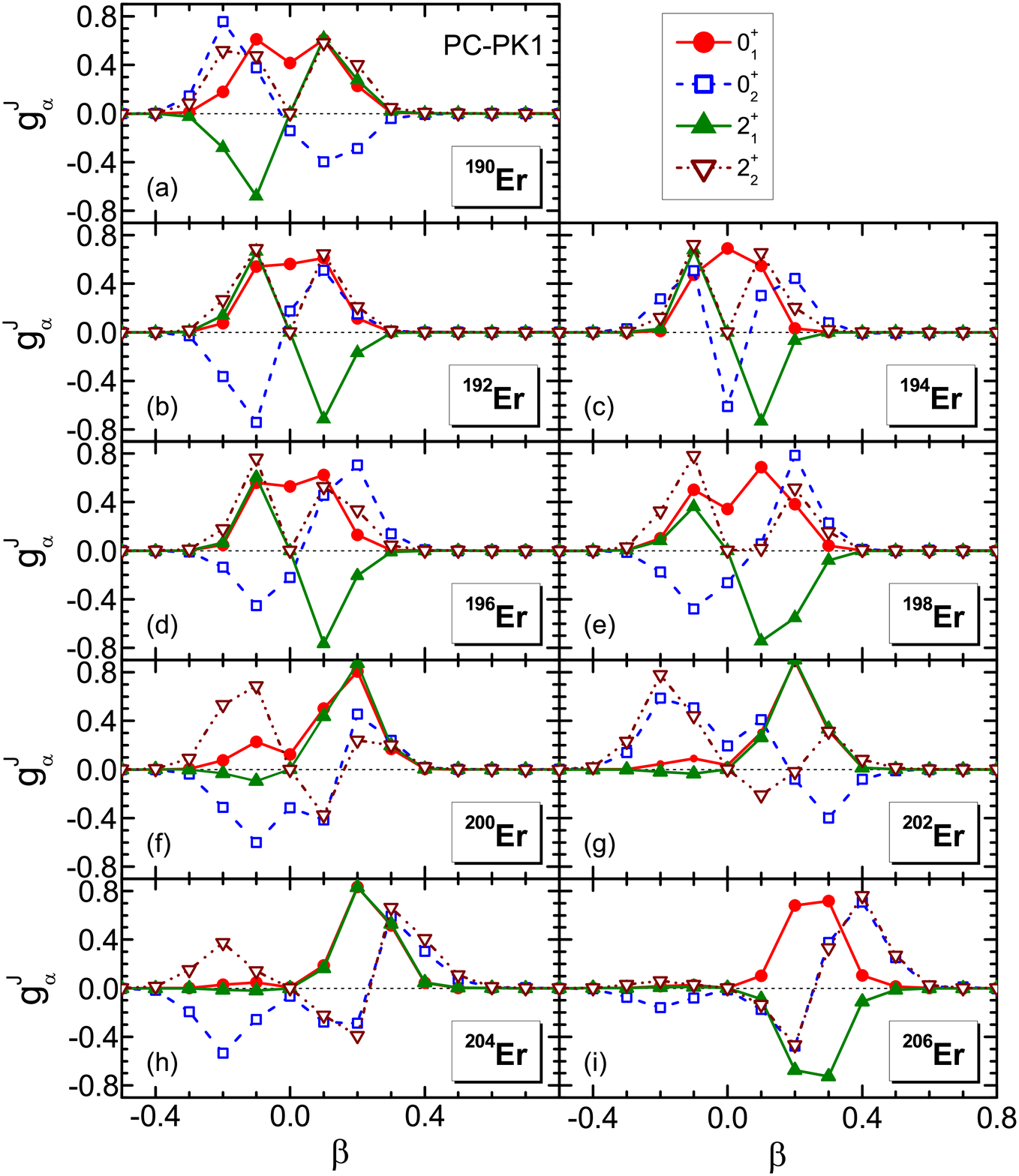}
 \caption{(Color online) Collective wave functions [cf. Eq.~\eqref{gfunc}] of the $0^+_1$, $0^+_2$, $2^+_1$, and $2^+_2$ states in $^{190-206}$Er. }
 \label{fig:wavfunEr}
 \end{figure}

 The evolution trend in the averaged quadrupole deformations of the first $0^+$ and $2^+$ states presents a clear picture of smooth shape transition with the increase of neutron number away from $N=126$. The second $0^+$ and $2^+$ states are more complicated, the averaged quadrupole deformations of which exhibit different behavior before and after $N=134$. 
 Since all the four isotopes share similar features, we only plot the collective wave functions for $^{190-206}$Er  in Fig.~\ref{fig:wavfunEr}.  It is seen that the  second $0^+$ and $2^+$ states  can be well approximated as spherical vibrational excitation states  in the isotopes with $N<134$. Beyond $N=134$, this structure is progressively destroyed by the increasing quadrupole deformation. The very sharp discontinuity around $N=134$ in the $0^+_2$ states might be interpreted as  a signature of spherical-to-prolate shape phase transition~\cite{Li09}.

 \begin{figure}[]
 \centering
 \includegraphics[width=8cm]{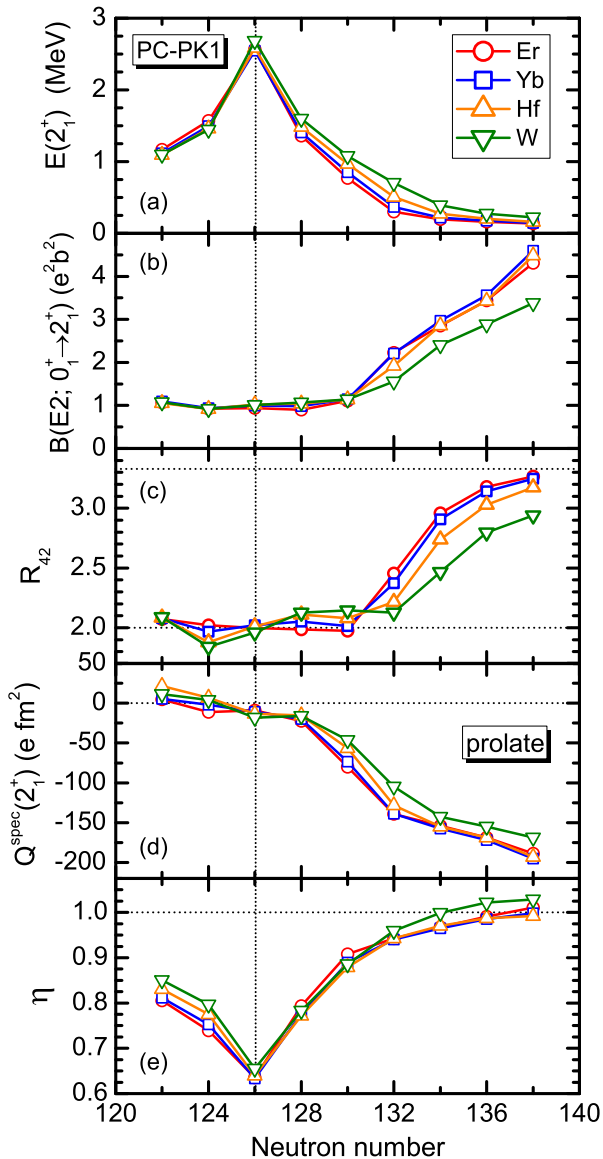}
 \caption{(Color online) (a) Excitation energy $2^+_1$ state, (b) electric quadrupole transition strength $B(E2; 0^+_1 \to 2^+_1)$, and (c) the  ratio of excitation energies $R_{42}=E(4^+_1)/E(2^+_1)$,  (d) spectroscopic quadrupole moment $Q^{{\rm spec}}(2^+_1)$, and (e) neutron-proton decoupling factor $\eta$ for Er, Yb, Hf, and W isotopes as a function of neutron number from the MR-CDFT calculations using PC-PK1. See text for details.}
 \label{fig:spectrum}
 \end{figure}

 Figure~\ref{fig:spectrum} displays the excitation energy of $2^+_1$ state, transition strength $B(E2; 0^+_1\to 2^+_1)$, the  ratio of excitation energies $R_{42}=E(4^+_1)/E(2^+_1)$, spectroscopic quadrupole moment $Q^{\texttt{spec}}(2^+_1)$, and neutron-proton decoupling factor $\eta$ as a function of neutron number.  The neutron-proton decoupling factor is defined as~\cite{Yao15,Wu15}
 \begin{equation}\label{eqeta}
 \eta=\frac{M_n/M_p}{N/Z} \, ,
 \end{equation}
 where $M_n$ and $M_p$ are the quadrupole transition matrix elements of neutrons and protons from ground state to $2^+_1$ state, respectively.
  A pronounced peak is found in the excitation energy of the $2^+_1$ state at $N=126$, which is consistent with the findings based on the Gogny force~\cite{Rodriguez15}, even though the value $\sim2.5$ MeV predicted in the present work is evidently smaller than their value $\sim4.5$ MeV.
 All the five  observables indicate the weakly quadrupole collectivity for the nuclei around $N=126$, which is consistent with the features of low-lying states in shell-closed nuclei. With the increase of neutron number, quadrupole collectivity is progressively developed with the predominate shape changing from spherical/weakly deformed one to prolate one when the neutron number is increased beyond $N\sim134$.
 The neutron-proton decoupling factor $\eta$ has a minimum at $N=126$. In other words, the quadrupole correlation contributed from neutrons  is much weaker than that from protons, indicating the magicity of the neutron number $N=126$.

 \begin{figure*}[]
 \centering
 \includegraphics[width=17cm]{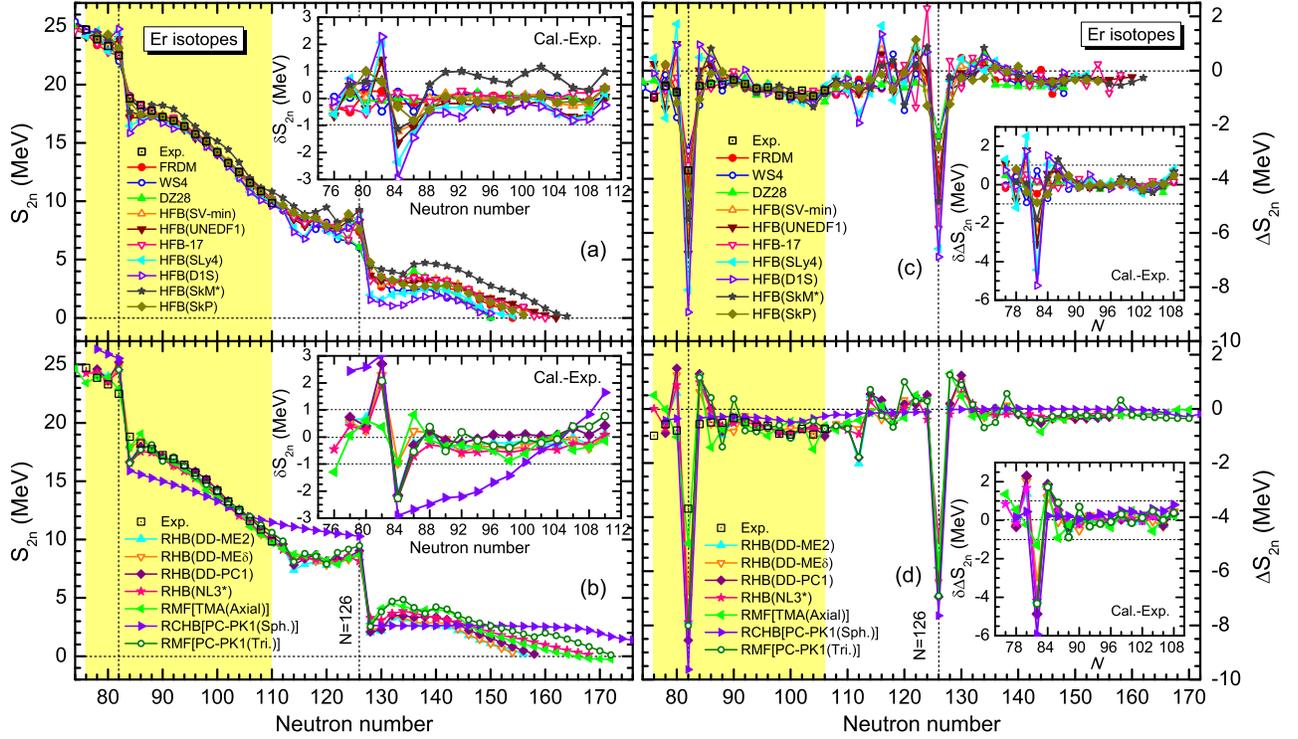}
 \caption{(Color online) Two-neutron separation energies (a, b) and their differentials (c, d) in Er isotopes as a function of neutron number. The upper panels (a, c) show the results from macroscopic-microscopic models and non-relativistic energy density functionals calculations, while the lower panels (b, d) are from the relativistic calculations. The discrepancy between model predictions and data is given in the insets. The data are taken from Ref.~\cite{Audi17}.   See text for details.}
 \label{fig:S2n}
 \end{figure*}

 \subsection{Analysis of the $N=126$ shell  closure}

We use two-neutron separation energies $S_{2n}$ and their differentials $\Delta S_{2n}(Z, N) = E(Z, N-2) + E(Z, N+2)-2E(Z, N)$, obtained from the masses of even-even nuclei, as a signature and measure of the underlying shell structure. Figure~\ref{fig:S2n} displays our predicted $S_{2n}$ and $\Delta S_{2n}$ in Er isotopes (labeled as RMF[PC-PK1(Tri.)]) as a function of neutron number across $N=82$ and $126$, in comparison with available data. Besides,  we perform a survey on separation energies obtained from other predictions, including 
\begin{itemize} 
\item macroscopic-microscopic (MM) mass models: the finite-range droplet model (FRDM)~\cite{Moller95}, the Weizs\"{a}cker-Skyrme mass model (WS4)~\cite{Wang14},  and the Duflo-Zuker mass model (DZ28)~\cite{Duflo95},

\item Hartree-Fock-Bogoliubov (HFB) models with non-relativistic Skyrme energy density functionals SV-min~\cite{Klupfel09}, UNEDF1~\cite{Kortelainen12}, HFB-17~\cite{Goriely09}, SLy4~\cite{Chabanat98},  SkM$^\star$~\cite{Bartel82}, SkP~\cite{Dobaczewski84},  and Gogny force D1S~\cite{Delaroche10},

\item as well as the relativistic mean-field (RMF) model \cite{Geng05a} with TMA~\cite{Sugahara94}, and relativistic Hartree-Bogoliubov (RHB) model~\cite{Afanasjev16, Afanasjev15,Agbemava14,Afanasjev13} with DD-ME2~\cite{Lalazissis05}, DD-PC1~\cite{Niksic08}, NL3$^\star$~\cite{Lalazissis09}, DD-ME$\delta$~\cite{Roca-Maza11}, and spherical relativistic continuum Hartree-Bogoliubov (RCHB) model~\cite{Xia18} with  PC-PK1~\cite{Zhao10}.  

\end{itemize}

Most of these mass tables are compiled in Ref.~\cite{maex}.   Before comparing the results by these models, some points should be kept in mind.  First of all, the parameters in the mass models (including HFB-17) are usually adjusted with all/most available data on nuclear masses. Therefore, the mass models generally show a better performance in the nuclei with data. They are not guaranteed to have the same performance on in-sample and out-of-sample nuclei. Second, even for the results from the calculations based on universal energy functionals, different types of approximation are used.  For the results labeled with ``HFB" or ``RHB/RCHB", pairing correlation between nucleons is treated with a general Bogoliubov transformation. Otherwise, the BCS approximation is used. Besides, for the results by energy functionals, axial symmetry is assumed if not specified.  The last but not least, different energy   functionals are fitted based on somewhat different protocols. It also introduces diversities into the predictions for neutron-rich nuclei.

The results by the non-relativistic  and  relativistic models are plotted in  Fig.~\ref{fig:S2n}(a)(c), and   Fig.~\ref{fig:S2n}(b)(d), respectively. One can see from the figure that
 
\begin{itemize}

\item All the models predict an abrupt drop in $S_{2n}$ and a deep peak in $\Delta S_{2n}$ at the magic numbers $N=82$ and $N=126$.
\item The discrepancy between the model predictions and available data for $S_{2n}$ and $\Delta S_{2n}$ is generally within 1.0 MeV, except for the nuclei around the shell closure $N=82$. 

\item The $S_{2n}$ of the nuclei with $N=82$ ($N=84$) is generally overestimated (underestimated) by all the energy  functionals. It brings an error up to $\sim3$ MeV into the predicted $\Delta S_{2n}$, in particular by the Skyrme SLy4, Gogny D1S and all the considered relativistic energy functionals except for TMA.

\item A very interesting finding is the steady increase in the  $S_{2n}$ starting from $N\sim 114$ up to $N=126$ predicted by most of the calculations.  The results of PC-PK1 with and without taking into account static deformation effect are significantly different from each other. Without the deformation effect, the $S_{2n}$ is decreasing monotonically with neutron number from $N=82$ to $N=126$ as shown in the results labeled with RCHB[PC-PK1(Sph.)]. It implies that the deformation effect is responsible for 
the  steady increase in the two-neutron separation energies before $N=126$. 

\end{itemize}

The above analysis is extended to Yb, Hf, and W isotopes and a similar phenomenon to that found in Er isotopes is observed, i.e., a large uncertainty exists in the predicted $S_{2n}$ and $\Delta S_{2n}$ by different models for the four isotopes around $N=126$.  The predicted $\Delta S_{2n}$ varies from $-2.0$ MeV to $-8.0$ MeV.

  \begin{figure}[]
 \centering
 \includegraphics[clip=,width=8.5cm]{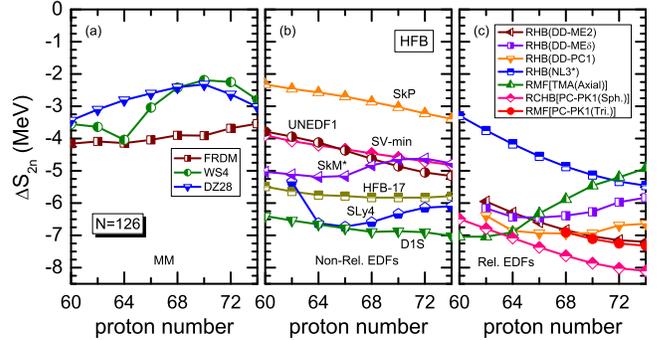}
 \caption{(Color online) The $\Delta S_{2N}$ obtained from different calculations as a function of proton number. }
 \label{fig:S2N126}
 \end{figure}

 Figure~\ref{fig:S2N126} displays the predicted $\Delta S_{2N}$ at $N=126$ from different models as a function of proton number from $Z=60$ to $Z=74$.
It is shown that the MM models predict a somewhat increase in the $\Delta S_{2n}$ with the decrease of proton number.
In contrast, the predicted $\Delta S_{2n}$ is quenching with the decrease of proton number by the non-relativistic Skyrme energy  functionals (except for the HFB-17, SLy4, and SkM$^\star$) and
the relativistic ones (except for the DD-ME$\delta$, DD-PC1, and TMA).  The evolution trend by  the TMA is obviously opposite to that by other relativistic functionals. It might be due to its mass-number dependent coupling strengths. Moreover,  we note that different  from other relativistic energy functionals that are usually optimized to a bunch of spherical nuclei in different mass regions, the DD-PC1 was optimized locally to 64 axially deformed nuclei in the mass regions $150\leq A \leq 180$ and $ 230\leq A \leq 250$.  The use of different fitting protocols may contribute partially to the  divergence in the predictions.

 \begin{figure}[]
 \centering
 \includegraphics[clip=,width=7.5cm]{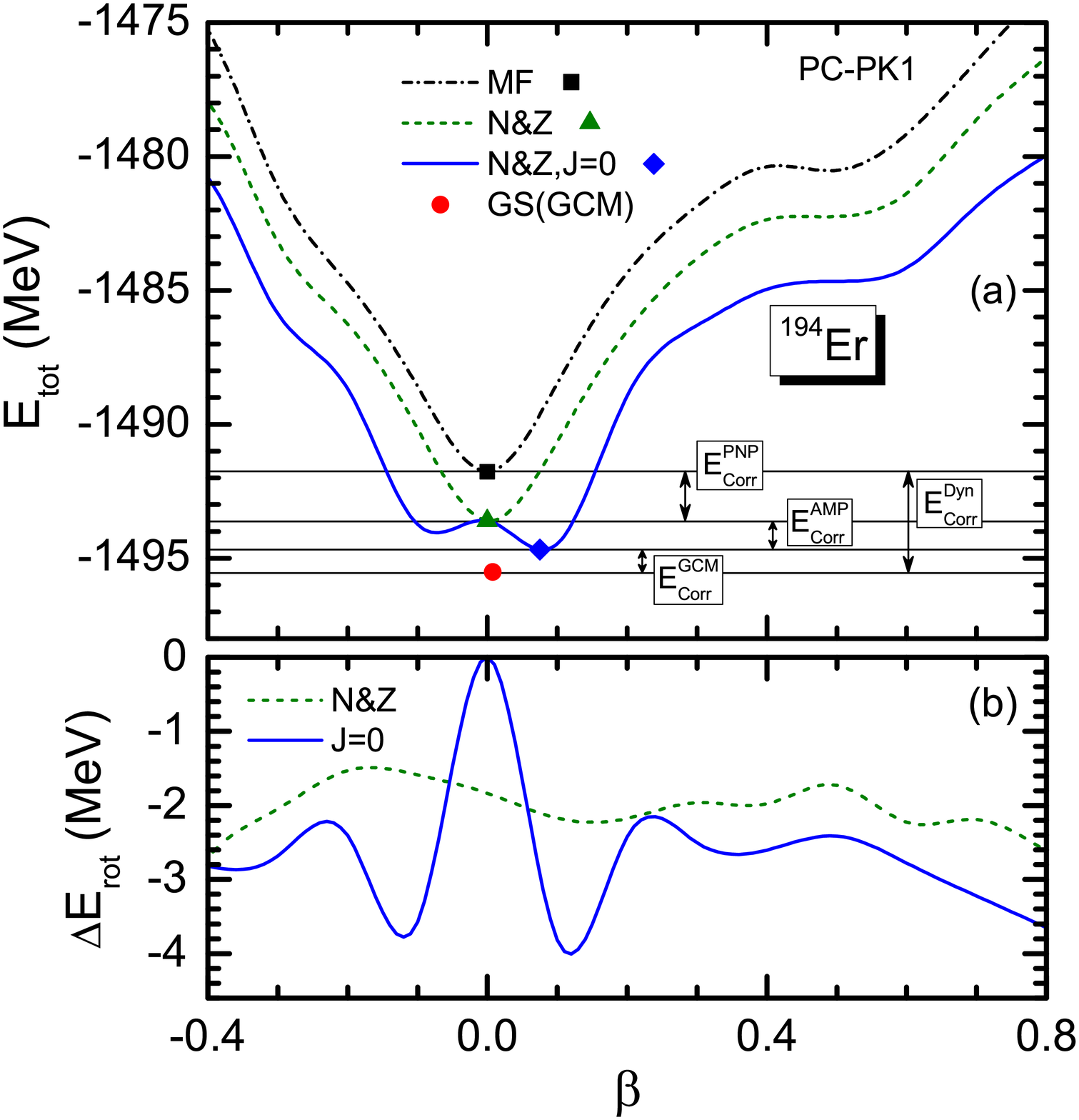}
  \caption{(Color online) (a) The energy surface of  mean-field states (MF), particle-number projected states ($N\&Z$),
           and those with an additional projection onto angular-momentum $J$ = 0 for $^{194}$Er by the relativistic PC-PK1.
           The  global energy minimum of each energy surface is marked with  square, triangle, and diamond symbols, respectively.
           The ground state by the GCM calculation is indicated with a red dot.
           (b) The energy gained from particle-number projection (dashed green line) and that from
           angular-momentum projection (solid blue line) as functions of the quadrupole deformation $\beta$. }
 \label{fig:PECsEr194}
 \end{figure}

 \begin{figure}[]
 \centering
 \includegraphics[width=8.5cm]{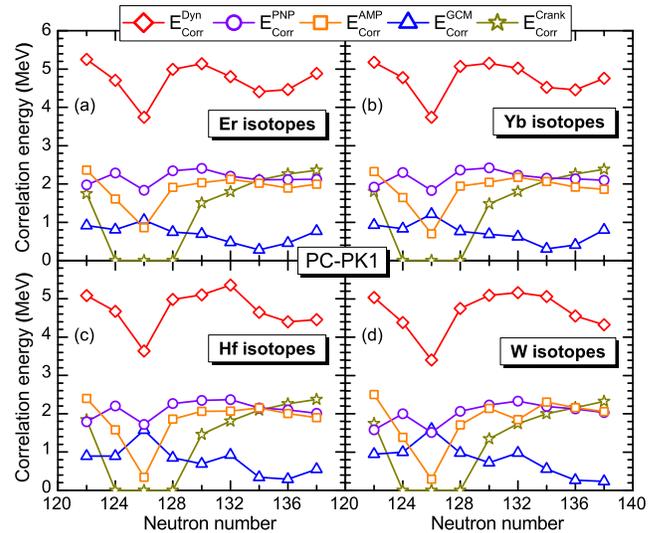}
 \caption{(Color online) Correlation energies $E_{\rm Corr}^{\rm Dyn}$, $E_{\rm Corr}^{\rm PNP}$, $E_{\rm Corr}^{\rm AMP}$, and $E_{\rm Corr}^{\rm GCM}$ for $^{190-206}$Er, $^{192-208}$Yb, $^{194-210}$Hf, and $^{196-212}$W isotopes as a function of neutron number.   The results $E^{\rm Crank}_{\rm Corr}$ calculated using the cranking prescription (\ref{dyn-cranking})  are also given for comparison.  }
 \label{fig:ECorr}
 \end{figure}

  \begin{figure}[]
 \centering
 \includegraphics[width=8cm]{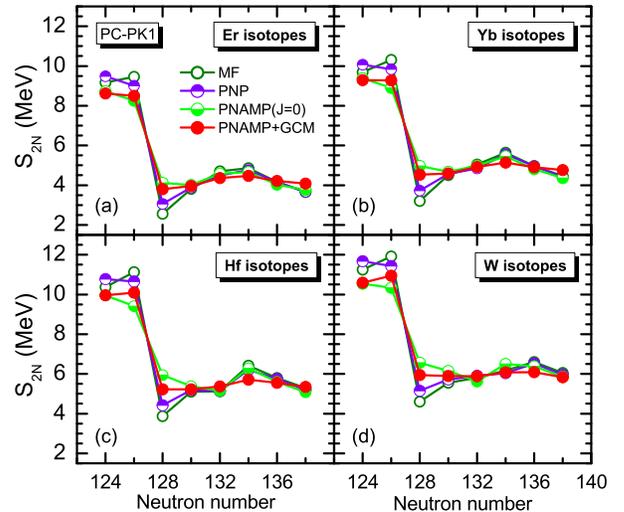}
 \caption{(Color online) Two-neutron separation energy from the MR-CDFT calculations with different approximation, including the pure mean-field calculation (MF),  with particle-number projection only (PNP) and projections onto both particle-number and angular momentum  $J = 0$ [PNAMP($J = 0$)], as well as the quantum-number projected GCM calculation (PNAMP+GCM),  for $^{190-206}$Er, $^{192-208}$Yb, $^{194-210}$Hf, and $^{196-212}$W isotopes as a function of neutron number.  }
 \label{fig:S2n-BMF1}
 \end{figure}

   \begin{figure}[]
 \centering
 \includegraphics[width=8cm]{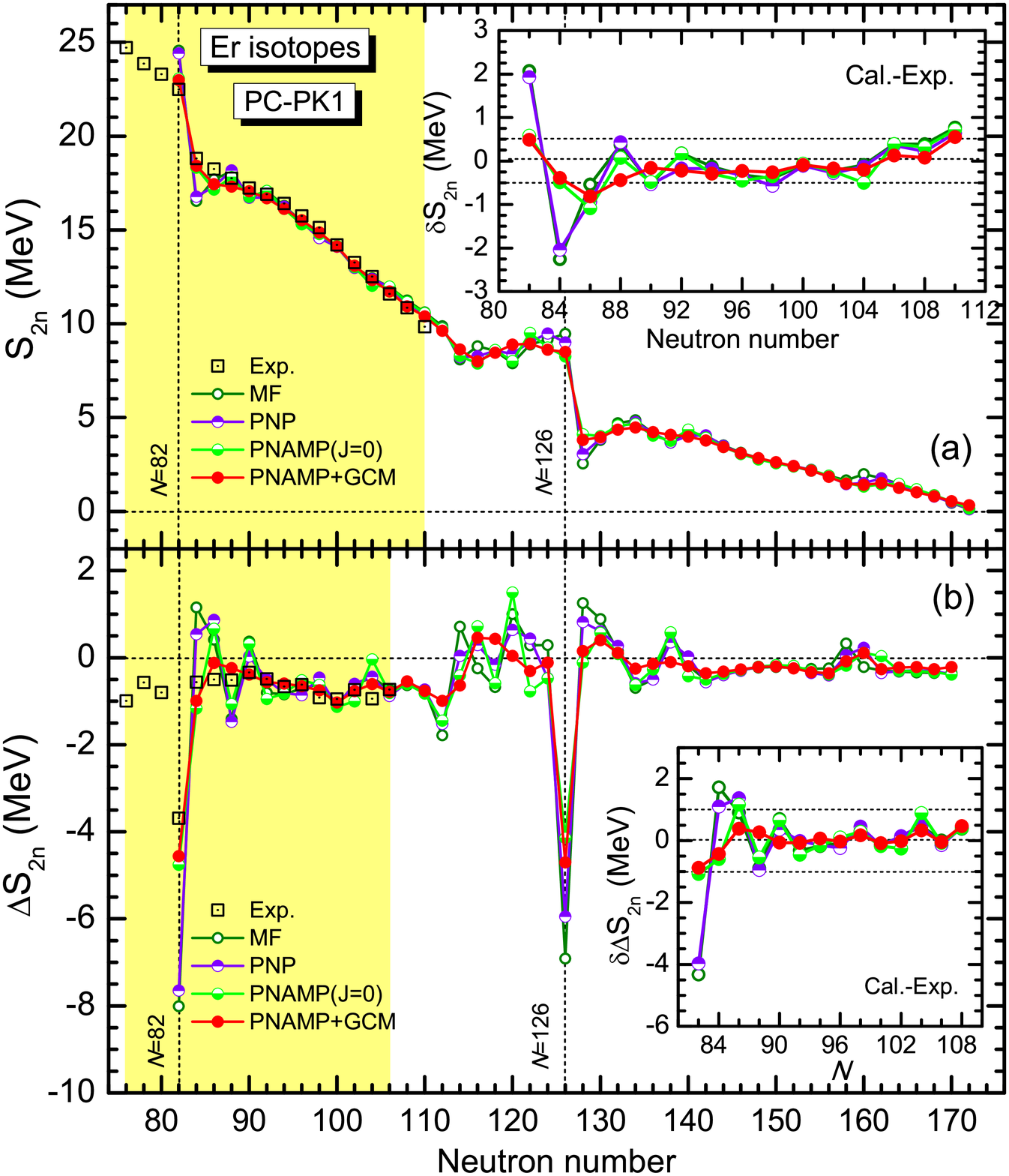}
 \caption{(Color online)  Two-neutron separation energies $S_{2n}$  (a) and their differentials $\Delta S_{2n}$ (b) in Er isotopes as functions of neutron number
  from the MR-CDFT calculations with different approximation using PC-PK1.}
 \label{fig:shell-gaps-Er}
 \end{figure}

  \begin{figure}[]
 \centering
 \includegraphics[clip=,width=8.0cm]{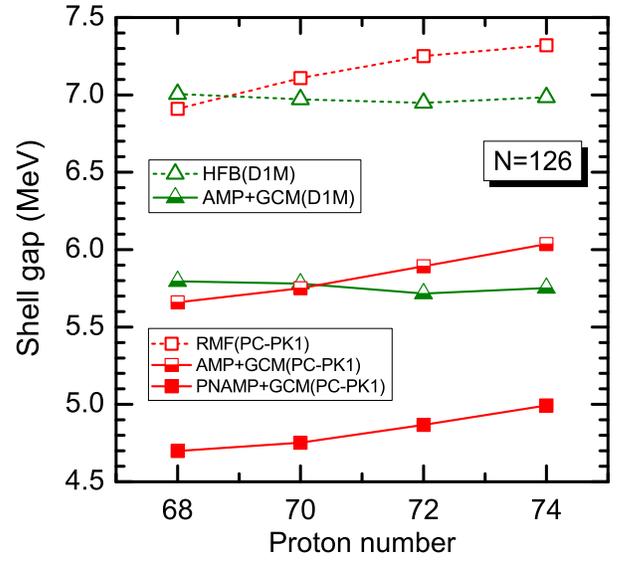}
 \caption{(Color online)  The absolute value of the differential $\Delta S_{2n}$ at neutron number $N=126$  from different calculations as a function of proton number. The results by the Gogny force D1M  are taken from Ref.~\cite{Rodriguez15}.  }
 \label{fig:S2Ngap}
 \end{figure}

 Subsequently, we examine the contribution of the BMF dynamic correlation energies to the predicted two-neutron separation energies. The dynamic correlation energy can be decomposed into three parts: the energy  $E_{\rm Corr}^{\rm PNP}$ from particle-number projection, the energy $ E_{\rm Corr}^{\rm AMP}$ from angular-momentum projection  and the energy  $E_{\rm Corr}^{\rm GCM}$ from shape mixing,
   \beqn
   \label{Ecorr}
  E^{\rm Dyn}_{\rm Corr} = E_{\rm Corr}^{\rm PNP} + E_{\rm Corr}^{\rm AMP} + E^{\rm GCM}_{\rm Corr}.
  \eeqn

  We take $^{194}$Er  as an example to illustrate each contribution. The results are shown in  Fig.~\ref{fig:PECsEr194}.  It is seen
  that about 2.07 MeV is gained in the energy from the particle-number projection, which shifts down the entire potential energy surface systematically.  The projection onto angular momentum $J=0$ brings additionally about 1.71 MeV contribution to the energy. The energy gained from shape mixing is around 0.98 MeV.  We carry out the same analysis for $^{190-206}$Er, $^{192-208}$Yb, $^{194-210}$Hf, and $^{196-212}$W. The results are displayed in Fig.~\ref{fig:ECorr}  as a function of neutron number. It is shown that the total dynamic correlation energy ranges between $\sim3.5$ MeV and $\sim5.5$ MeV, with the minimum located at $N=126$.  For comparison, we also evaluate the dynamic correlation energy using the cranking prescription~\cite{Goriely05,Cham08,Goriely09,Zhang14}
  \begin{equation}
  \label{dyn-cranking}
  E^{\rm Crank}_{\rm Dyn}=E_{\rm rot}\{b{\rm tanh}(c|\beta_m|)+d|\beta_m|{\rm e}^{-l(|\beta_m|-\beta^{0})^{2}}\}\, ,
  \end{equation}
 where the values of parameters $b, c, d, l,$ and $\beta^0$ are chosen as 0.80, 10, 2.6, 10, and 0.10 according to Ref.~\cite{Cham08}. The $E_{\rm rot}$ represents the rotational correction to the energy
 \begin{equation}
 \label{rot}
 E_{\rm rot}=\frac{\hbar^{2}}{2{\cal I}}\langle\hat{J}^{2}\rangle,
 \end{equation}
 where the moment of inertia ${\cal I}$ is calculated by the Inglis-Belyaev formula and $\hat{J}$ corresponds to the angular-momentum operator. As expected, the phenomenological formula (\ref{dyn-cranking}) underestimates systematically the dynamic correlation energy, in particular for the
 nuclei around $N=126$.  It means that the cranking prescription cannot be used for the purpose of study dynamic correlation effects on the neutron separation energies of nuclei around shell closure.

 \begin{figure}[]
 \centering
 \includegraphics[width=8cm]{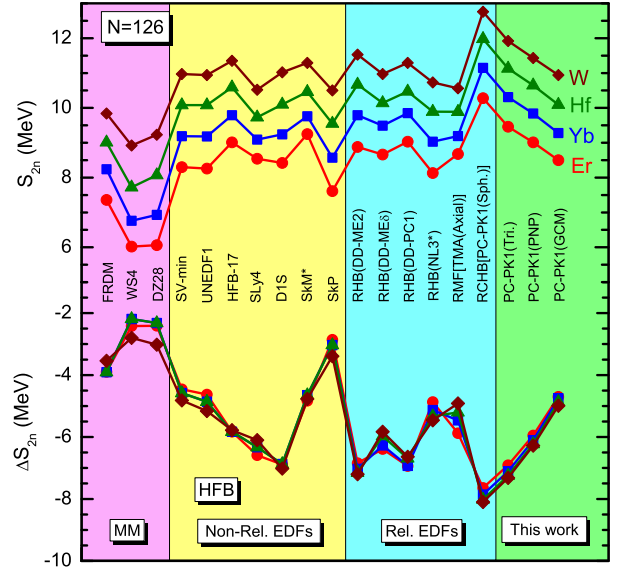}
 \caption{(Color online) Two-neutron separation energies (upper) and their differentials (lower) at neutron number $N=126$ in Er, Yb, Hf, and W isotopes predicted by different models. }
 \label{fig:shell-gaps-models}
 \end{figure}

Since the amount of dynamic correlation energy varies from nucleus to nucleus, it affects the predicted two-nucleon separation energy.  The previous studies for stable nuclei have demonstrated that the two-neutron separation energies are overall  improved after taking into account the dynamic correlation energies~\cite{Zhang14,Lu15,Wu15,Bender05,Bender06,Tomas15}. In Fig.~\ref{fig:S2n-BMF1}, we show how the dynamic correlation energy evaluated at different level changes the two-neutron separation energies in the four isotopes.  In the mean-field results, an increase in the two-neutron separation energy with neutron number ranging from $N=128$ to $N=134$ is attributed to deformation effect,   which was discussed in Ref.~\cite{Rodriguez15}.  With the inclusion of dynamic correlations from particle-number projection and angular-momentum projection, the amount of energy dropping in $S_{2n}$ from $N=126$ to $N=128$ is dramatically decreased. This effect becomes moderate after taking into account shape mixing in the GCM calculation, in which case, the variation of the two-neutron separation energy with neutron number is much smoother.

We extend the above analysis to the entire Er isotopes. The separation energies $S_{2n}$ together with their differentials $\Delta S_{2n}$ are shown in Fig.~\ref{fig:shell-gaps-Er}. 
The BMF effects quench the variation of $S_{2n}$ with neutron number, in particular around $N=82$ and $N=126$. In the region with available data around $N=82$,
one can see that these effects reduce significantly the discrepancy  between theoretical results and data.

Figure~\ref{fig:S2Ngap} shows the predicted differential $\Delta S_{2n}$ of two-neutron separation energy at $N=126$ as a function of proton number, in comparison with the results from the calculations using the Gogny force D1M~\cite{Rodriguez15}. The $N = 126$ shell gap is quenched when the BMF correlations are included. It is shown that  angular-momentum projection decreases the $\Delta S_{2n}$ by about 1.3 MeV in both cases. This value is further decreased by $\sim1.0$ MeV with  the particle-number projection. Besides, we note that  the $\Delta S_{2n}$  by AMP+GCM with D1M is almost a constant (about 5.8 MeV) with the decrease of proton number from $Z=74$ to $Z=68$, while  that by the PC-PK1 is decreasing evidently from $\sim5.0$(6.0) MeV to $\sim4.7$(5.7) MeV.  The origin of this difference is not clear yet, but might be related to the different isospin dependent spin-orbit potential.

 Figure \ref{fig:shell-gaps-models} summarizes the two-neutron separation energy and its differential at $N=126$ predicted by different models. Generally, the predicted absolute value of $\Delta S_{2n}$ by the MM models (FRDM, WS4, and DZ28) is overall smaller than those by the energy  functional calculations (except for the  Skyrme force SkP \cite{Dobaczewski84}).  On the mean-field level, the Gogny force D1S and relativistic energy functionals DD-ME2 and PC-PK1 predict the largest values for the $N=126$ shell gap. For the PC-PK1,  the BMF effects reduce the $\Delta S_{2n}$ to be around $-5.0$ MeV, quenching the shell gap by $\sim 30\%$.
 
 To asses the possible impact of $\sim 30\%$ shell quenching at $N=126$ by the PC-PK1 on the $r$-process abundances in a qualitative way, we compare the BMF effects on the  $S_{2n}$  and $\Delta S_{2n}$ 
 from our calculation with that from the 5DCH calculation based on the Gogny D1S force \cite{Delaroche10} in Table \ref{s2nD1SPCPK1}. The latter has been adopted into the $r$-process calculations by Arcones and Bertsch.  It was found that the BMF effects on the masses reduce significantly the trough in the abundances before the third peak at $A\sim 195$~\cite{Arcones12}, which are similar to the shell quenching effects on the $r$ process~\cite{Chen95,Arcones11}. 
One can see from Table \ref{s2nD1SPCPK1} that the BMF effects by the 5DCH based on D1S  decrease the $S_{2n}$ and $\Delta S_{2n}$ much more pronounced than that by the GCM based on PC-PK1. The quenching effect at $N=126$ shell gap in the former is larger than the latter by about a factor of two.  It indicates a much more moderate influence on the $r$ process  abundances from the BMF effects by the GCM calculation using PC-PK1 than that illustrated in Ref~\cite{Arcones12}. Of course, a dedicated $r$-process calculation with the entire mass table by the PC-PK1 is required  before drawing a solid conclusion.

 \begin{table}[t]
 \caption{Two-neutron separation energy $S_{2n}$ and its differential $\Delta S_{2n}$ at $N=126$ from both mean-field and BMF calculations based on either relativistic
 PC-PK1 or Gogny D1S. The BMF results of D1S were evaluated with the 5DCH from Ref.~\cite{Delaroche10}. See text for  details.}
 \begin{ruledtabular}
 \begin{tabular}{ccccccccccccccccccccc}
   & $Z$ & \multicolumn{4}{c}{$S_{2n}$ (MeV) } & \multicolumn{4}{c}{$\Delta S_{2n}$ (MeV)}     \\
 \cline{3-6\  \ }  \cline{7-10\  \ }
  & & \multicolumn{2}{c}{PC-PK1}   &\multicolumn{2}{c}{D1S}  & \multicolumn{2}{c}{PC-PK1}   &\multicolumn{2}{c}{D1S}   \\
 \cline{3-4\  \ } \cline{5-6\  \ }  \cline{7-8\  \ } \cline{9-10\  \ }
 &  &  \multicolumn{1}{c}{RMF}   &\multicolumn{1}{c}{GCM}   & \multicolumn{1}{c}{HFB} & \multicolumn{1}{c}{5DCH} &  \multicolumn{1}{c}{RMF}   &\multicolumn{1}{c}{GCM}   & \multicolumn{1}{c}{HFB} & \multicolumn{1}{c}{5DCH} \\
 \hline
 &$68$     & 9.46    & 8.50   & 8.42   & 5.81  & -6.91    & -4.70  &-6.89   & -3.07   \\
 &$70$     & 10.31   & 9.28   & 9.24   & 6.64  & -7.11    & -4.75  &-6.87   & -3.00   \\

 &$72$    & 11.13   & 10.09  & 10.10  & 7.48  & -7.25    & -4.87  &-6.90   & -2.90   \\
 &$74$    & 11.92   & 10.94  & 11.02  & 8.31  & -7.32    & -4.99  & -7.02  & -2.72   \\

 \end{tabular}
 \end{ruledtabular}
 \label{s2nD1SPCPK1}
 \end{table}
 
%
\section{Summary}\label{Sec.V}
%

We have presented a comprehensive study of neutron-rich Er, Yb, Hf, and W isotopes across the $N=126$ shell  with the MR-CDFT. With the techniques of quantum-number projections and GCM, we have calculated the observables of low-lying states  using the PC-PK1 parameterization of the relativistic point-coupling Lagrangian density. 
Our results have shown that  the quadrupole collectivity is progressively developed  in all the four isotopes with the increase of neutron number beyond $N=126$.  It corresponds to a  transition from spherical shape to prolate deformed one.  The general features of the low-lying states in closed-shell nuclei, i.e., a sharp peak of $2^+_1$ excitation energies and a pronounced neutron-proton decoupling,  have been found in the isotopes around $N=126$, indicating the robustness of $N=126$ shell gap. 

Besides,  we have studied the impact of BMF effects on the predicted nuclear masses, two-neutron separation energies and their differentials.  The
BMF effects quench the variation of the two-neutron separation energies  with neutron number and lead to a reduction in the predicted shell gaps. 
For the $N=126$ shell gap, this quenching effect is $\sim 30\%$  in all the four isotopes, consistent with that found in the GCM calculation based on the Gogny forces~\cite{Tomas15,Rodriguez15}, but much smaller than that in the 5DCH calculation~\cite{Delaroche10}.  It implies that the
BMF effect on the $r$ process through nuclear masses will be more moderate  than that found in Ref.~\cite{Arcones12}, in which the BMF effects are evaluated with the 5DCH calculation~\cite{Delaroche10}.  A quantitative investigation of these effects on the $r$-process abundances is required before  drawing a solid conclusion. This kind of investigation
asks for a global mass-table calculation with the MR-CDFT.   Work along this direction is in progress.
 

\begin{acknowledgements}
 The authors thank S. Giuliani for fruitful discussion and careful reading of the manuscript. This work was supported in part by the National Natural Science Foundation of China
under Grants No. 11575148 and No. 11475140, the Jiangxi Normal University (JXNU) Initial Research Foundation Grant to Doctor (12019504), and the Young Talents Program under JXNU. JMY acknowledges the support from the National Science Foundation under Grant No. PHY-1614130, as well as the U.S. Department of Energy, Office of Science, Office of Nuclear Physics under Grants No. de-sc0017887 and de-sc0018083 (NUCLEI SciDAC Collaboration). 
\end{acknowledgements}


\end{document}